\documentstyle{europhys}

\def\stars{\bigskip\centerline{***}\medskip}

\newif\ifboo \boofalse

\begin{document}
\euro{}{}{}{}
\Date{}
\shorttitle{K. ELDER {\it et al.} AN X-RAY SCATTERING ETC}
\title{ An X-ray Scattering and Simulation Study
of the Ordering Kinetics in CuAu}
\author{K. R. Elder\inst{1}, Oana Malis\inst{2}, Karl Ludwig\inst{2},
Bulbul Chakraborty\inst{3} and Nigel Goldenfeld\inst{4}}
\institute{
\inst{1} Department of Physics, Oakland University, Rochester, MI, 48309-4401 \\
\inst{2} Department of Physics Department, Boston University,
590 Commonwealth Ave.,
Boston, MA
02215 \\
\inst{3} Physics Department, Brandeis University, Waltham, MA, 02254 \\
\inst{4} Department of Physics, University of Illinois at Urbana-Champaign,
1110 West Green Street, Urbana, Illinois 61801 }

\rec{}{}

\pacs{
\Pacs{05}{70Ln}{Nonequilibrium thermodynamics}
\Pacs{64}{60Cn}{Order/disorder transformations}
\Pacs{81}{30Hd}{Constant-composition solid/solid phase transformations}
}
\maketitle
\begin{abstract}
A detailed numerical and experimental study of the ordering of the
low temperature tetragonal phase of CuAu
is presented.  The numerical simulations are based on a
coarse-grained free energy derived from electronic
structure calculations of CuAu, while the experimental results
are obtained from {\it in situ} x-ray scattering.
Both theoretical and experimental work indicate a
subtle kinetic competition between the ordered tetragonal
phase and the metastable modulated phase.
\end{abstract}

What determines the microstructure of a typical alloy?
While there is no simple answer to this question it is
understood\cite{gms83,b94} that a variety of morphological 
instabilities can be encountered in non-equilibrium processing 
which create spatial inhomogeneities that are magnified during
coarsening.  In this paper experimental and theoretical techniques 
are combined to investigate the intricate morphologies that emerge 
during the sublattice ordering of the low temperature tetragonal 
phase of CuAu.  The results of this study indicate a subtle 
competition between a metastable modulated phase and the stable 
low temperature ordered phase.  At shallow quench temperatures or 
early times the modulated state is favored while at lower temperature 
or later times the ordered phase dominates.  The results are
supported by {\it in situ} x-ray scattering experiments and
numerical simulations. 

	The phases of CuAu that are relevant to this study 
are a high temperature disordered face-centered cubic phase,
an intermediate temperature modulated orthorhombic phase and
a low temperature ordered tetragonal phase.  The modulated phase
consists of a periodic array of antiphase domain walls arranged 
along one of the directions perpendicular to the tetragonal direction.
The wavelength of the modulated phase is 10 underlying fcc unit cells. 
The transitions from the ordered phase to the modulated phase and from 
the modulated phase to the disordered phase are both first order 
and occur at $T_{OM}=658K$ and $T_{MD}=683K$ respectively.  
Rapid temperature quenches through these transitions 
can lead to many interesting morphologies.  In this paper  
quenches to the low temperature ordered phase (i.e., $T<T_{OM}$) 
are considered.  Earlier experiments\cite{tuhy93} 
suggested homogeneous nucleation for quenches just below $T_{OM}$ 
and spinodal ordering for deep quenches. 

	In principle the ordering kinetics involve both 
structural relaxation of the lattice and a local rearrangement of 
atoms into the appropriate sublattice.  Although it is desirable 
to use a microscopic model that describes these two distinct 
processes\cite{EMT1,cx92}, it is difficult to explore the length and 
time scales of interest in this paper with such an
atomistic approach.  To study these long length 
and time scales a mean-field continuum model\cite{cx92,ceg95} 
will be used.  The mean-field model is based on a microscopic 
Hamiltonian derived from effective medium theory(EMT)\cite{EMT1}.  
EMT is a semi-empirical approach based on the concept of
an electron-density dependent contribution to the total energy of a
system of atoms.  The parameters of this model were fit to 
experimental values of the lattice parameter, bulk
modulus and shear modulus of Cu and Au and to the heat of
formation of the 50-50 alloy\cite{EMT1}.  The Hamiltonian describing 
Cu-Au alloys through the whole concentration range is defined by 
the EMT functional form and these parameters. The mean-field model 
is obtained from this Hamiltonian and describes the free energy of 
the system as a functional of the sublattice concentration and
explicitly assumes that the structural relaxation is instantaneous.
The model predicts the existence of the phases and transitions described 
above. As with most mean-field theories the predictions for the 
transition temperatures (i.e., $T_{MD}=1263K$ and $T_{OM}=1211K$) are not 
in good agreement with the experimental values\cite{foot2}. An enhanced
mean-field theory could be constructed to improve the
quantitative details, but the competition between ordered and
modulated phases predicted by the current model already agrees 
qualitatively with experiment.

	Assuming that the kinetics are driven by minimization of 
this free energy the following kinetic model can be proposed,
\begin{equation}
\label{eq:model}
\partial \eta /\partial t = - 2 \Gamma(2\pi/a)^d
([a_T+e(\nabla^2_\perp-\nabla^2_{||})+f\nabla_\perp^4]\eta 
-4u\eta^3+6v\eta^5) + \zeta, 
\end{equation}
where, 
$\left<\zeta(\vec{r},t) \zeta(\vec{r},t) \right> = -2k_B T \Gamma 
\delta(\vec{r}-\vec{r}')\delta(t-t')$, $a_T = 0.042(T-1219K) meV/K$ 
and $\eta$ represents the sublattice concentration.
All parameters, except $\Gamma$, the mobility, were determined by 
the EMT calculations described above and are given in 
previous work\cite{cx92}.  
Rather than introducing a phenomenological or empirical form for $\Gamma$ 
all times will be recorded in terms of a Ginzburg-Landau time defined such 
that the fastest growing Fourier mode grows as $e^{t/t_{GL}}$. For this 
model $t_{GL} = 1/[2\Gamma(2\pi/a)^d(e^2/4f-a_T)]$.
The subscripts $\perp$ and $||$ refer to directions perpendicular and 
parallel to the ordering directions respectively.  The interesting structures
develop in the plane perpendicular to the ordering direction 
and for computational efficiency gradients in the $||$ direction will 
be ignored.  This model differs from the standard kinetic model 
of sublattice ordering (Model A) since the coefficient of the 
$\nabla^2_{\perp}$ term is negative.  Eq. (\ref{eq:model}) was 
simulated on a discrete 
lattice using Euler's method to evaluate the time derivatives 
and a `spherical Laplacian' introduced by Oono and Puri\cite{op87} 
to account for the spatial gradients.  Relatively small space 
and time steps and large system sizes were used to eliminate 
spurious numerical effects.  Eq. (\ref{eq:model}) was simulated 
on system of dimension $2048 \times 2048$ grid points corresponding 
to a physical dimension of $0.5\mu m \times 0.5\mu m$.

The three equilibrium phases, disordered, modulated and ordered are
described by $\left<\eta\right> = 0$, 
$\eta \approx A\sin(q_o\vec{r}\cdot\hat{n})$ 
and $\left<\eta\right>=\eta_o$ respectively, where $\hat{n}$ is 
an arbitrary unit vector in the ordering plane and 
$q_o$ is the modulation wavevector.  This differs 
from CuAu for which the modulation wavevector appears only 
along the Bravais lattice directions in the ordering plane.
Transmission electron microscope work\cite{owwk58,gp59} has indeed 
shown that the morphologies in the modulated regime are 
quite anistropic.  Nevertheless, even at this level of simplification 
the model contains many complexities that are important to 
the ordering process. Most importantly it incorporates the 
subtle competition between the modulated, disordered and ordered 
phases.

	To examine the kinetics of ordering below $T_{OM}$
several simulated quenches were examined.  In the first set of 
quenches the system was equilibrated in the disordered 
phase (i.e., at $T-T_{MD}=20K$) and then instantaneously 
quenched to temperatures below $T_{OM}$.  A qualitative 
picture of the subsequent dynamics is shown in fig. (\ref{config1})
for quench temperatures of $T_{OM}-T = 5K$ and $15K$.  
At both temperatures the very early stages are dominated by
highly interconnected structures that contain many defects.
The subsequent kinetics are dominated by a coarsening in 
which the average length scale of the patterns increases.
This coarsening occurs mainly near defects in the pattern. 
Figure (\ref{config1}) also indicates that the lower quenches coarsen more 
rapidly.  This is due to the increase in the free energy difference 
between the modulated and ordered states at lower 
temperatures.  Although not shown here this effect was 
verified at other quench temperatures (i.e., $T-T_{OM}=13K$ and $65K$).

	To further illustrate the kinetics of ordering and 
for comparison with the experimental x-ray scattering patterns 
it is useful to consider the spherically 
averaged dynamic structure factor, $S(q,t)$.  $S(q,t)$ is defined as,
\begin{equation}
S(q,t) \equiv \sum_{q^2=q_x^2+q_y^2} |\eta(\vec{q},t)|^2 /
\sum_{q^2=q_x^2+q_y^2} 1,
\end{equation}
where $\eta(\vec{q},t)$ is the discrete Fourier transform of 
$\eta(\vec{x},t)$.  The ordered and modulated phases are represented by a 
peak(s) in $S(q,t)$ centered around $q_O=0$ and $q_M=\pm .18 /\AA$ 
respectively.  To examine the competition between the two phases 
the structure factor was fit to two symmetric peaks: an ordered peak,
$S_O(k,t)$,  and a modulated peak, $S_M(k,t)$, centered 
around $q_O$ and $q_M$ respectively.  For comparison with 
experiment the integrated intensities (i.e., 
$\int d\vec{q} S_O(q,t)$ and $\int d\vec{q} S_M(q,t)$) were 
evaluated for each peak.  Ideally, these quantities reflect the volume 
fraction of the two phases. In practice there is no clear distinction 
between `ordered' and `modulated' regimes as is illustrated 
in morphological patterns shown in fig. (\ref{config1}).  Nevertheless 
these quantities are useful for comparison with experiment as they 
are relatively insensitive to experimental resolution (which is not 
the case for peak heights and widths).  The dynamics of the 
integrated intensities and some representative structure 
factors are shown in fig. (\ref{theory}).

	The structure factors in fig. (\ref{theory}) indicate
that satellite peaks are generated at early times for 
quenches just below $T_{OM}$ consistent with the highly 
interconnected patterns shown in fig. (\ref{config1}a) and 
(\ref{config1}c). The metastability of the modulated phase 
is highlighted by the persistence of the satellite peaks at late 
times as seen in fig. (\ref{theory}a) and (\ref{theory}b).  
At the lower quench temperature (see figs. (\ref{theory}c) and 
(\ref{theory}d)) the central peak associated with the ordered 
phase dominates more rapidly.  

	To test the numerical predictions and to provide 
additional insight into the ordering, x-ray scattering 
studies on the modulated and ordered superlattice peaks 
near (110) were carried out. The experiments were conducted 
on beamline X20C at the National Synchrotron Light Source.  
Several kinds of CuAu samples were studied -- including
bulk single crystals and polycrystalline films approximately
$10\mu m$ thick. The detailed ordering kinetics differed
slightly between samples but the qualitative features have been 
reproduced consistently. The results presented in this paper 
are from the polycrystalline films. 
The samples were held in a $He$ atmosphere and {\it in situ} x-ray 
diffraction patterns were recorded by a linear position-sensitive 
photodiode array detector every $3s$.  Quench rates were 
typically of the order $5.5$ degrees/s.  Similar to the numerical 
simulation the prequench samples were equilibrated at $T-T_{MD}=20K$
and then rapidly quenched to temperatures below $T_{OM}$.  
All x-ray 
intensities, $I(q,t)$, were normalized with the incident intensity 
and a linear function was subtracted from each pattern to remove the 
background. 

	The x-ray scattering results for quenches to $T_{OM}-T= 5K$ 
and $15K$ are summarized in fig. (\ref{experiment}) and can 
be compared directly with the numerical results shown in 
fig. (\ref{theory}).  In this comparison it should be noted that 
the numerical results are for correlations in the sublattice 
concentration ($\eta$) which, for example, peak at $q=0$ for the 
ordered phase (i.e., $\eta=\eta_o$).  In the experiment 
the same ordered phase leads to a superlattice peak or Bragg reflection 
centered around $2\pi/d$ where $d$ is the (110) plane spacing. 
Both experiment and simulation show that satellite peaks are generated 
at early times and that the persistence of these peaks is related to the 
quench temperature.  Although not shown here, it was also found, 
in both experiment and simulation, that the early stage kinetics are 
very similar above and below $T_{OM}$.

One feature that was observed in the experiment and not in 
the simulations was the shift in position with time of the x-ray
peaks.  This motion arises from changes in both the 
lattice spacing and modulation wavelength.  
The fundamental (200) peaks (not presented here) directly show the evolution 
of the lattice structure\cite{KIN}. 
Examination of these indicates that, at the temperatures
and time scales 
discussed below, the lattice has already relaxed to its tetragonal shape. 
Another significant effect is the shifting of the satellite peaks away 
from the central superlattice peak.   These changes are 
typically quite large and occur in the early to intermediate 
time scales. For example in the quench to $T_{OM}-T=15K$ the 
modulation wavelength decreased from approximately 15.6 to 11.5 
lattice constants on a time scale of $300s$. 

Diffuse scattering studies of the disordered alloy have found
satellites around the superlattice peaks with a ``modulation'' wavelength 
approximately 40\% larger than in the equilibrium modulated state\cite{DIFF}.
In the early stages of ordering following a quench, these
satellite peaks grow rapidly in height before moving outward
to the equilibrium positions.  

The ordering kinetics can be strongly altered by changing 
the prequench state from a disordered phase to a modulated phase 
since, in the model,
the modulated phase is metastable at low temperatures and 
the disordered phase is unstable.  To examine this 
phenomenon two deep quenches were considered, one from a disordered 
initial condition (as before) and one from a modulated state.
The numerical prequench state was created by seeding the system 
with large modulated regimes and evolving the system 
until no disordered patches remained.  The resulting pattern 
was a modulated state containing many defects.   For comparison the 
prequench experimental system was equilibrated in the modulated phase at 
$T-T_{OM} = 3K$.  

Quenching these states to $T_{OM}-T=65K$ led to the results 
illustrated in  fig. (\ref{metastab}).  The influence of the 
prequench state can be clearly seen in figs. (\ref{metastab}a) 
and (\ref{metastab}b) which compare configurations at the 
same time following the quench.  The persistence of the prequench 
modulation is quite apparent in fig. (\ref{metastab}b).  
The erosion of this modulated phase occurs only at defects in 
the pattern.  In principle the ordered regions can 
spontaneously nucleate in a purely modulated regime, however it 
appears that the nucleation rates are very small.  For example, in 
one test simulation conducted at $T_{OM}-T=65K$ no nucleation 
events were observed for times upto $t/t_{GL}=5,000$.  This 
is also apparent although less obvious in the quenches described 
earlier (see fig. (\ref{config1})).

	The experimental and theoretical results for 
these quenches are compared in figs. (\ref{metastab}c) and 
(\ref{metastab}d).  In fig. (\ref{metastab}c) the scattering 
intensity is shown at $1650$ seconds after the quench for 
both prequench states.  A similar comparison is provided in 
fig. (\ref{metastab}d) for the numerical simulations.  
In both experiment and simulation growth of the ordered state 
is seen to be strongly reduced for the modulated prequench state.
This can be seen by comparing the amplitudes of the central and 
satellite peaks for the two different prequench states.

	In summary, the combined experimental and theoretical 
work presented in this paper has provided a detailed description 
of the kinetics of ordering in the low temperature tetragonal 
phase of CuAu.  This description highlights the influence of 
the metastable modulated phase on the ordering process 
and the interesting kinetics that result.  The continuum model 
provided a good qualitative prediction of the competition between 
the ordered and modulated phases as a function of time and 
pre- and post-quench temperatures. 

\stars

The authors would like to thank Bill Klein and Nick Gross for many 
useful discussions.  The work of KE was supported by a Research
Corporation Grant \#CC4181.  
The work of BC was supported by NSF grants  
DMR-9208084 and DMR-952093.  NG acknowledges support from NSF 
though grant DMR-93-14938. The Boston University component of this 
research was supported by NSF under DMR-9633596.  The NSLS is supported 
by DOE Division of Materials Sciences and Division of Chemical Sciences.

\eject

\begin{figure}
\caption{Transient morphologies for $T_{OM}-T=5K$
and $15K$ obtained from numerical simulations.
The light and dark regions respectively correspond to
$\eta < 0$ and $\eta > 0$.  Only one twenty fifth of the
full simulations cell ($0.5\mu m \times 0.5\mu m$)
is shown here.  Figures (a) and (b) correspond to
$T_{OM}-T=5K$ at times $t/t_{GL}= 112$ and $375$
respectively.
Figures (c) and (d) depict $\eta$ for
$T_{OM}-T=15K$ at times $t/t_{GL}= 49$ and $146$
respectively.}
\label{config1}
\end{figure}

\begin{figure}
\caption{In figures (a) and (c) the numerical
simulated integrated intensities
are shown as a function of time for quenches to
$T_{OM}-T=5K$ and $T_{OM}-T=15K$ respectively.
The solid, long-dashed and short-dashed lines
correspond to the modulated, ordered and total
integrated intensities respectively.  In figures
(b) and (d) sample structure factors are shown
at $T_{OM}-T=5K$ and $T_{OM}-T=15K$ respectively
at the times indicated by the arrows in (a) and (c).
The lines from bottom to top at $q=0$ correspond to
the earliest to latest times.}
\label{theory}
\end{figure}

\begin{figure}
\caption{In figures (a) and (c) the experimentally
measured integrated intensities
are shown as a function of time for quenches to
$T_{OM}-T=5K$ and $T_{OM}-T=15K$ respectively.
The solid, long-dashed and short-dashed lines
correspond to the modulated, ordered and total
integrated intensities respectively.  In figures
(b) and (d) sample scattering patterns are shown
at $T_{OM}-T=5K$ and $T_{OM}-T=15K$ respectively
at the times indicated by the arrows in (a) and (c).
}
\label{experiment}
\end{figure}

\begin{figure}
\caption{In figs. (a) and (b) configurations are shown
for at $t/t_{GL} = 18.8$ for quenches from
disordered and partially modulated phases respectively.
As with fig. (1) only one twenty fifth of the full simulation
cell is shown.  In fig. (c) the experimental scattering intensity
is shown at $1650s$ after the quench for both disordered (dashed line)
and modulated (solid) initial states.  A similar comparison is
shown for the numerical structure factor in (d) at $t/t_{GL}=18.8$
after the quench.}
\label{metastab}
\end{figure}

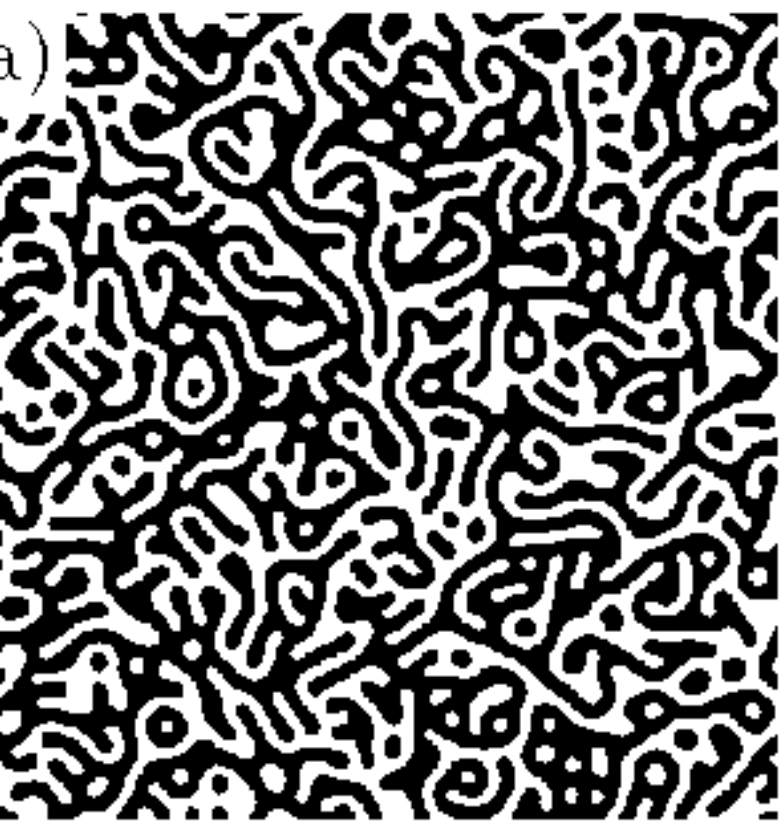
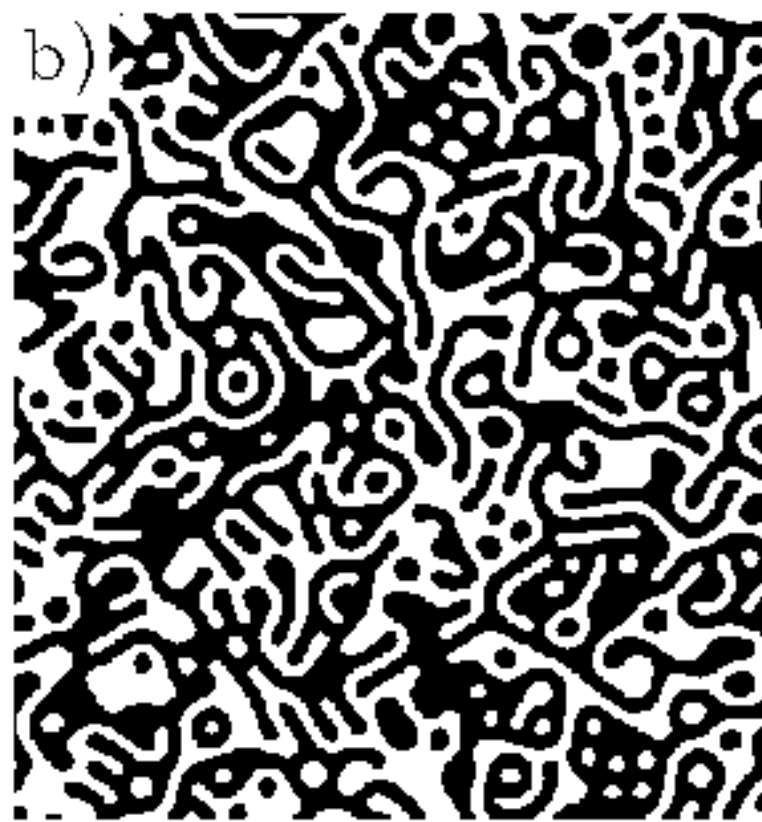
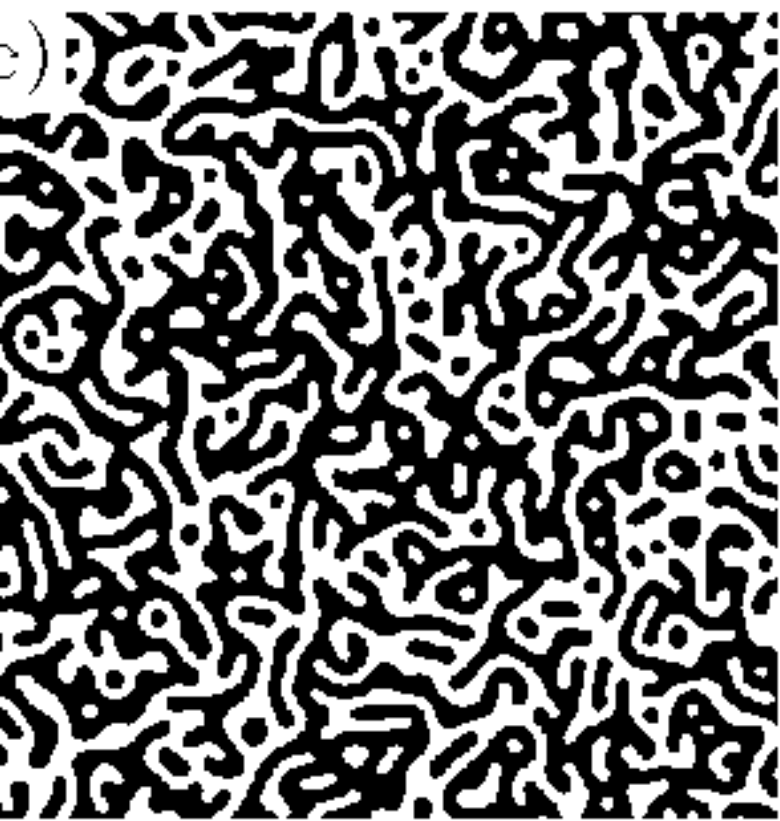
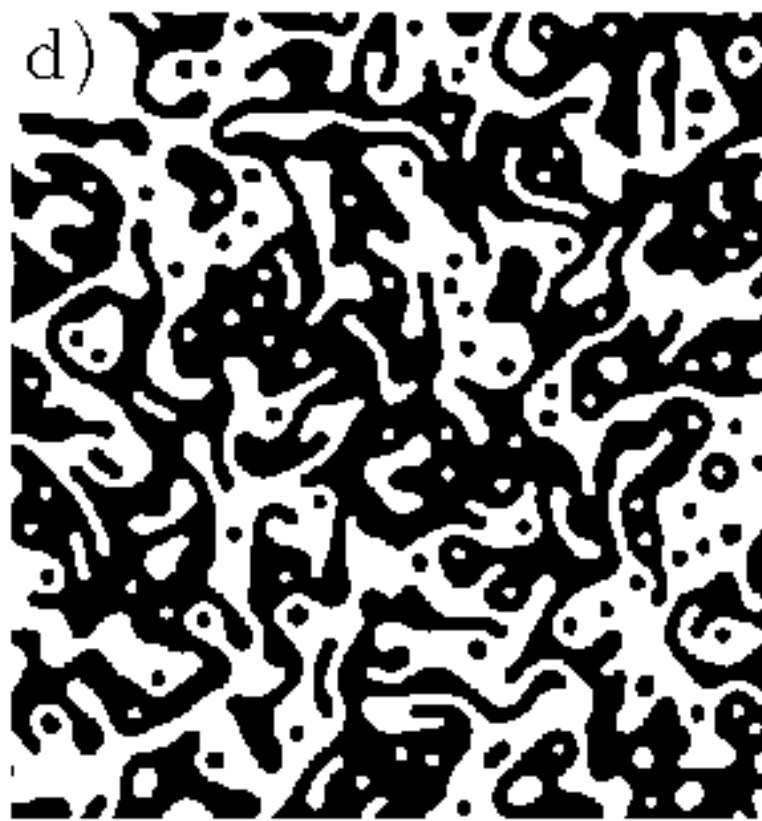
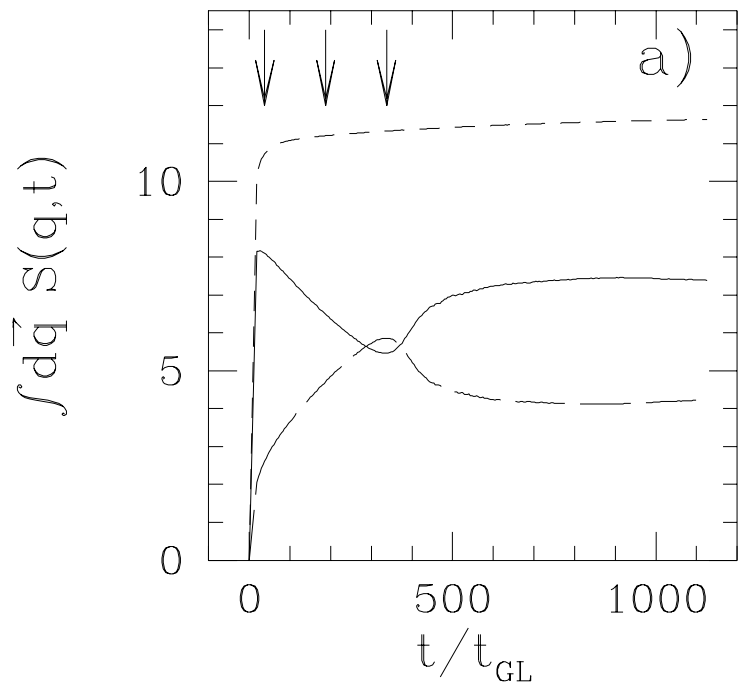
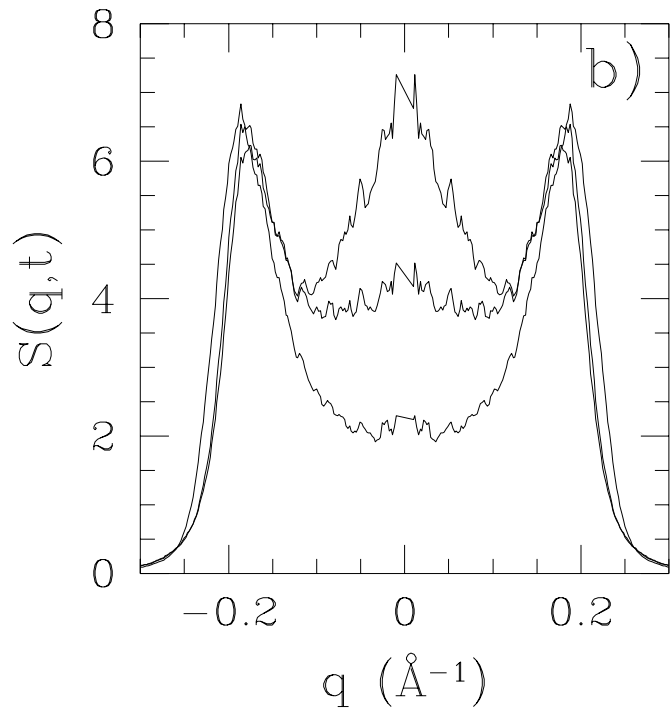
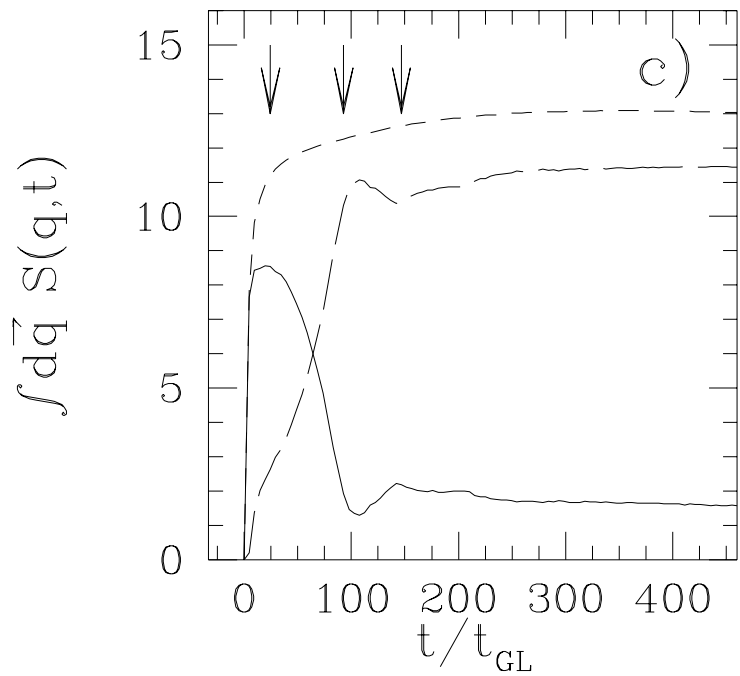
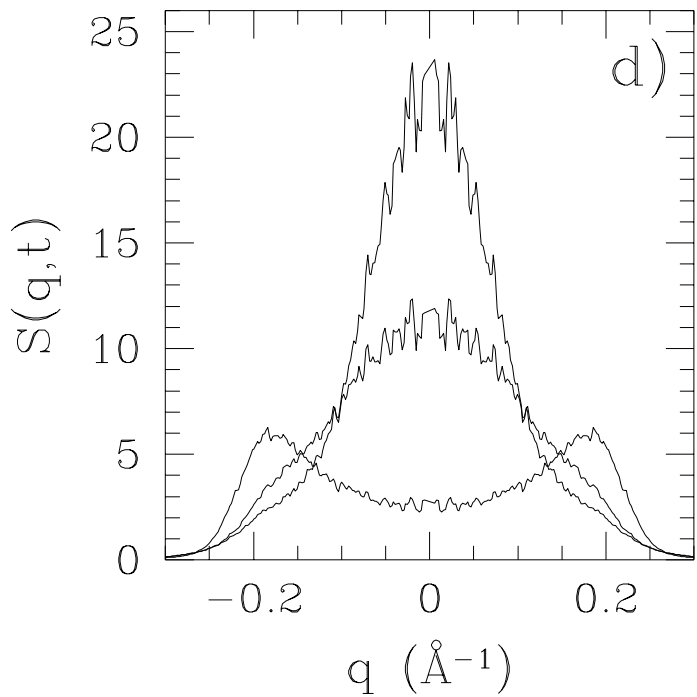
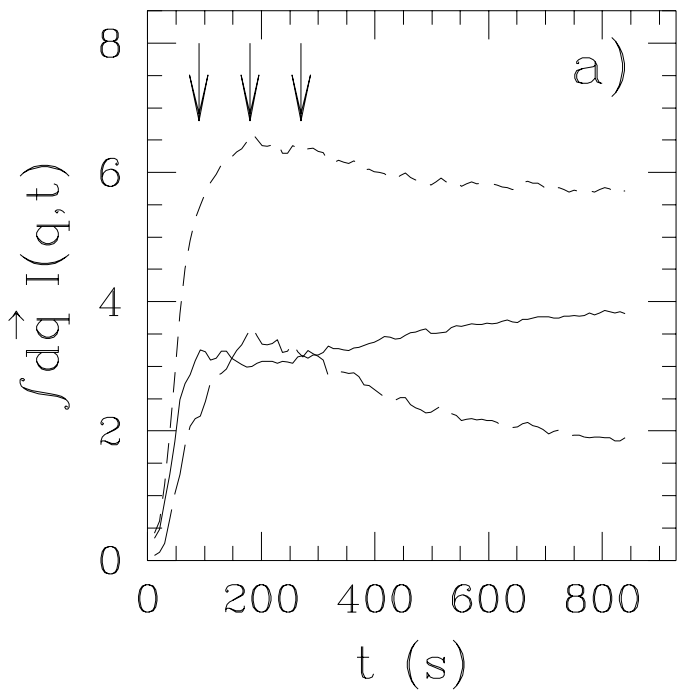
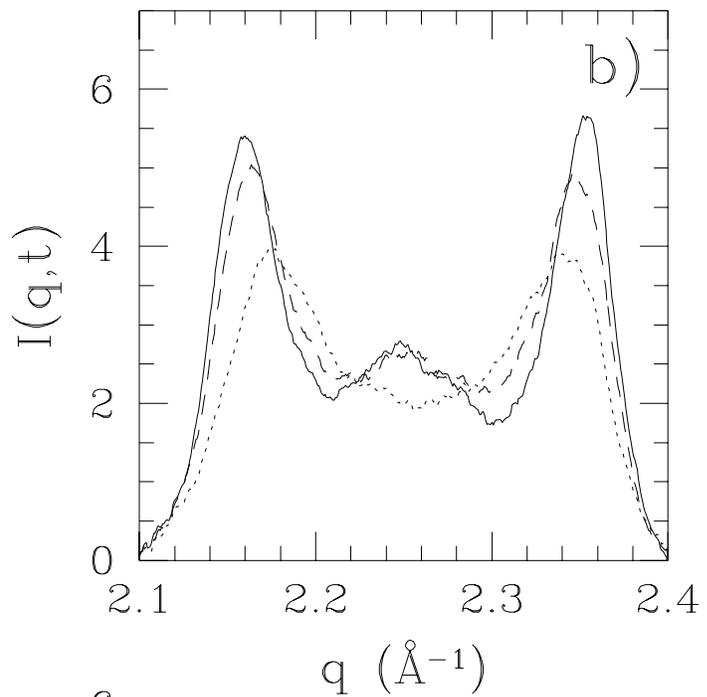
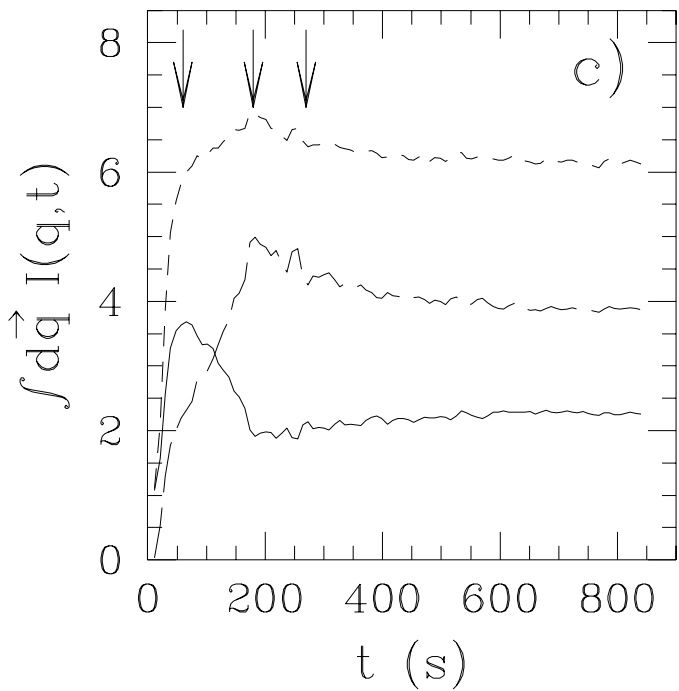
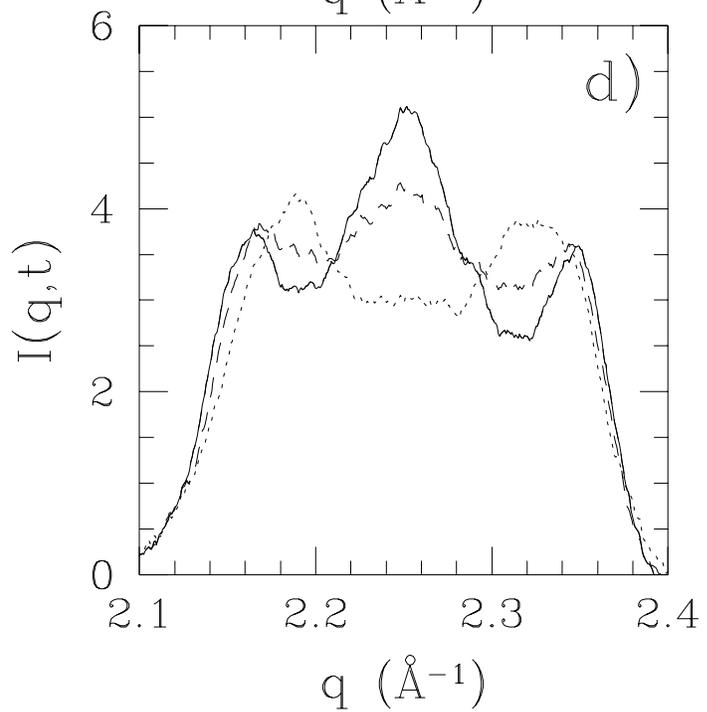
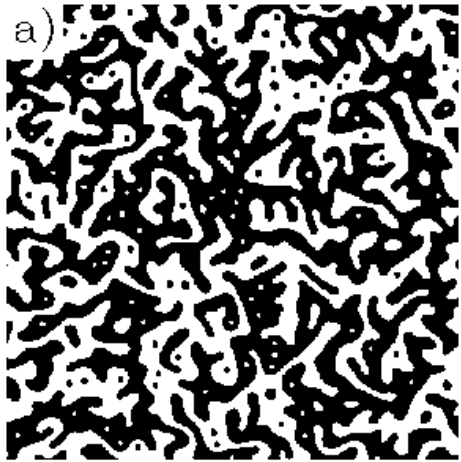
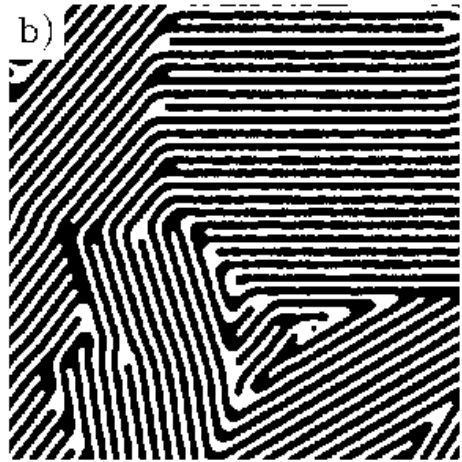
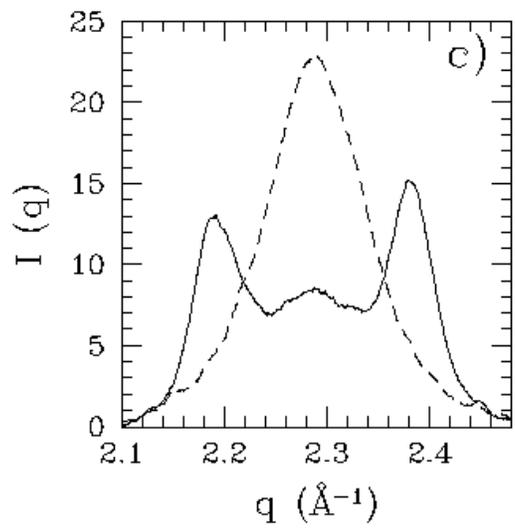
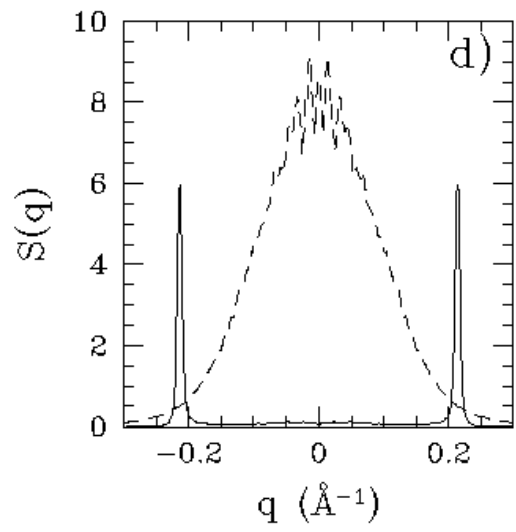
\end{document}